\documentclass[journal]{IEEEtran}

\makeatletter
\long\def\@makecaption#1#2{\ifx\@captype\@IEEEtablestring%
\footnotesize\begin{center}{\normalfont\footnotesize #1}\\
{\normalfont\footnotesize\scshape #2}\end{center}%
\@IEEEtablecaptionsepspace
\else
\@IEEEfigurecaptionsepspace
\setbox\@tempboxa\hbox{\normalfont\footnotesize {#1.}~~ #2}%
\ifdim \wd\@tempboxa >\hsize%
\setbox\@tempboxa\hbox{\normalfont\footnotesize {#1.}~~ }%
\parbox[t]{\hsize}{\normalfont\footnotesize \noindent\unhbox\@tempboxa#2}%
\else
\hbox to\hsize{\normalfont\footnotesize\hfil\box\@tempboxa\hfil}\fi\fi}
\makeatother

\usepackage{graphicx}
\usepackage{amssymb,amsmath,bm}
\usepackage{multirow}
\usepackage{cite}
\usepackage{hhline}

\begin{document}

\title{Bridging the Gap Between Monaural Speech Enhancement and Recognition with Distortion-Independent Acoustic Modeling}

\author{Peidong~Wang,~\IEEEmembership{Student Member,~IEEE}, \\
		Ke~Tan,~\IEEEmembership{Student Member,~IEEE},
        and~DeLiang~Wang,~\IEEEmembership{Fellow,~IEEE}
        \thanks{This research was supported in part by two NSF grants (IIS-1409431 and ECCS-1808932) and the Ohio Supercomputer Center.}
        \thanks{P. Wang is with the Department of Computer Science and Engineering, The Ohio State University, Columbus, OH 43210 USA  (e-mail: wang.7642@osu.edu).}
        \thanks{K. Tan is with the Department of Computer Science and Engineering, The Ohio State University, Columbus, OH 43210 USA (e-mail: tan.650@osu.edu).}
        \thanks{D. L. Wang is with the Department of Computer Science and Engineering and the Center for Cognitive and Brain Sciences, The Ohio State University, Columbus, OH 43210 USA (e-mail: dwang@cse.ohio-state.edu).}
}
\maketitle

\begin{abstract}
Monaural speech enhancement has made dramatic advances since the introduction of deep learning a few years ago. Although enhanced speech has been demonstrated to have better intelligibility and quality for human listeners, feeding it directly to automatic speech recognition (ASR) systems trained with noisy speech has not produced expected improvements in ASR performance. The lack of an enhancement benefit on recognition, or the gap between monaural speech enhancement and recognition, is often attributed to speech distortions introduced in the enhancement process. In this study, we analyze the distortion problem, compare different acoustic models, and investigate a distortion-independent training scheme for monaural speech recognition. Experimental results suggest that distortion-independent acoustic modeling is able to overcome the distortion problem. Such an acoustic model can also work with speech enhancement models different from the one used during training. Moreover, the models investigated in this paper outperform the previous best system on the CHiME-2 corpus.
\end{abstract}

\IEEEpeerreviewmaketitle

\section{Introduction}
\IEEEPARstart{F}{ormulated} as a supervised learning problem, speech enhancement has made major progress over the last few years with the use of data driven methods, particularly deep learning. Wang and Wang \cite{wang2012boosting,wang2012cocktail} first introduced deep neural networks (DNNs) to perform time-frequency (T-F) masking for speech enhancement. Lu \emph{et al.} and Xu \emph{et al.} used a deep autoencoder (DAE) or DNN to map from the power spectrum of noisy speech to that of clean speech \cite{lu2013speech,xu2014experimental,xu2015regression}. Many subsequent studies have been conducted to perform T-F masking or spectral mapping by employing a variety of deep learning models, acoustic features, and training targets \cite{weninger2015speech,nie2014deep,xu2017multi,xu2014dynamic,gao2015improving,li2015dnn}. These studies have elevated the performance of speech enhancement by a large margin \cite{wang2018supervised}. DNN-based monaural speech enhancement has improved, for the first time, the intelligibility of noisy speech for human listeners with hearing impairment as well as those with normal hearing \cite{healy2013algorithm,wang2018supervised,kolbk2017speech}.

Along with the progress in speech enhancement, researchers have investigated using speech enhancement models as frontends for automatic speech recognition (ASR) systems. Narayanan \emph{et al.} \cite{narayanan2013ideal,narayanan2014investigation} proposed to combine masking-based DNN speech enhancement with speech recognition. With a Gaussian mixture model (GMM) as backend, the enhancement frontend was shown to reduce word error rate (WER) significantly \cite{narayanan2013ideal}. In a subsequent paper using DNN as backend, the benefit of speech enhancement is mixed, depending on training features \cite{narayanan2014investigation}. For the acoustic model trained with cepstral features, speech enhancement still helps. With log-Mel features, however, the enhancement frontend causes performance degradation. Du \emph{et al.} \cite{du2014robust} applied mapping-based frontends to both GMM and DNN based recognition backends. Their observations are basically in line with those of Narayanan \emph{et al.} The only difference is that their enhancement frontend can yield improvements on clean, noisy, and clean plus channel-mismatched conditions for the DNN acoustic model trained with noisy speech. In the fourth CHiME speech separation and recognition challenge (CHiME-4), Heymann \emph{et al.} \cite{jahn2016wide} noted that the harm of processing artifacts introduced during enhancement may outweigh the benefit brought by noise reduction. Based on these studies as well as our own attempts in applying monaural speech enhancement as a frontend for speech recognition on CHiME-4 corpus, the distortion to speech signals introduced in monaural speech enhancement is a major problem that can render enhancement useless or even harmful for robust ASR.

One way to alleviate the distortion problem is to reduce or eliminate speech distortions in enhancement frontends. Attempts in this direction include a progressive training scheme proposed by Gao \emph{et al.} \cite{gao2016snr} and a mimic loss proposed by Bagchi \emph{et al.} \cite{bagchi2018spectral}. Progressive training \cite{gao2016snr} fine-tunes enhancement models in a multitask manner. Instead of using clean speech as the only target of output layer, they add multiple layers in DNN treating speech with progressively decreased signal-to-noise ratio (SNR) as labels. This way, the enhancement model is trained to reduce noise gradually, as well as the distortion in output layer. The mimic loss based method \cite{bagchi2018spectral,plantinga2018exploration} jointly trains enhancement frontends and recognition backends. It uses senone labels directly as the training target. Experimental results showed that such enhancement frontends can be used with off-the-shelf ASR models in Kaldi \cite{povey2011kaldi} on the second CHiME speech separation and recognition corpus (CHiME-2).

In addition to pursuing distortion reduction in speech enhancement models, designing more distortion tolerant acoustic model backends may be another direction. Previous research in speech enhancement field shows that DNNs trained using a variety of noises have the ability to generalize to new noisy conditions \cite{chen2016large}. A recent study performed by Narayanan \emph{et al.} \cite{narayanan2018toward} investigated the generalization ability of acoustic models trained with various out-of-domain data (noises, bandwidths, codecs, and features). Their observation is that, through large-scale training, such acoustic models perform as well as acoustic models trained with in-domain data.

In this study, we analyze the distortion problem by viewing it as a noise mismatch between training and testing. After comparing five acoustic models, we find that distortion-independent acoustic model can potentially overcome the distortion problem. Experimental results also show that this type of acoustic model can work with speech enhancement frontends different from the one used during training.

The rest of this paper is organized as follows. Section II gives an analysis of the distortion problem, an explanation of distortion-independent acoustic modeling, and a description of utterance-wise recurrent dropout for acoustic model training. Sections III and IV present the experiment setup and results, respectively. We make concluding remarks in Section V.

\section{System Description} \label{system_description}
\subsection{An Analysis on the Distortion Problem}
The distortion in this study refers to the alteration to clean speech signal introduced by speech enhancement that may cause performance degradation in an ASR system. More specifically, this paper tackles with the distortion problem of noise-independent speech enhancement. The input to a speech enhancement system is generated by mixing clean speech with an additive noise, as shown below:

\begin{equation}
\label{eq:exp_time}
y = s + n
\end{equation}
where $y$ denotes noisy speech, $s$ clean speech, and $n$ an additive noise.

The frequency domain representation of Eq. (\ref{eq:exp_time}) can be written as (\ref{eq:exp_freq}) below:

\begin{equation}
\label{eq:exp_freq}
Y = S + N
\end{equation}
where $Y$, $S$, and $N$ are the spectral representations of noisy speech, clean speech, and additive noise, respectively.

Speech enhancement typically operates on the magnitudes of frequency domain representations. Masking-based models generate a T-F mask, which is then element-wise multiplied with the magnitude of $Y$,

\begin{equation}
\label{eq:masking}
|\hat{S}| = |Y| \otimes M = |S+N| \otimes M
\end{equation}
where $|\cdot|$ denotes magnitude, $\otimes$ element-wise multiplication, $\hat{S}$ enhanced speech, and $M$ mask.

Depending on the T-F mask definition, $M$ is typically a real-valued matrix with element values ranging from zero to one, e.g. the ideal ratio mask (IRM) \cite{wang2014training}. For such masks, (\ref{eq:masking}) can be written as below:

\begin{equation}
\label{eq:masking2}
|\hat{S}| = |S + N| \otimes M = |S \otimes M + N \otimes M|
\end{equation}

Thus, we have

\begin{equation}
\label{eq:masking3}
|\hat{S}| = | S + S \otimes (M - A) + N \otimes M|
\end{equation}
where $A$ is an all-one matrix.

The distortion for ASR backends can be defined as:

\begin{equation}
\label{eq:distortion}
D = S \otimes (M - A) + N \otimes M = N \otimes M - S \otimes \overline{M}
\end{equation}
where $\overline{M}$ denotes the complement of $M$.

There are two special cases of $D$. First, if $M$ is an all-one matrix, speech enhancement has no impact on noisy speech. The influence of $S$ on $D$ can also be ignored. Second, let us consider the case when $M$ equals the IRM defined below:

\begin{equation}
\label{eq:irm}
IRM = \frac{|S|}{|S| + |N|}
\end{equation}

In this case, $D$ will be an all-zero matrix, and the distortion problem does not exist.

Other than the two cases above, the influence of $S$ cannot be ignored and the second term in (\ref{eq:distortion}) can be viewed as noise residue, which is different from $N$. Due to this residue, distortion is different from noise $N$.

\begin{figure}

\begin{minipage}[b]{1.0\linewidth}
  \centering
  \centerline{\includegraphics[width=6cm]{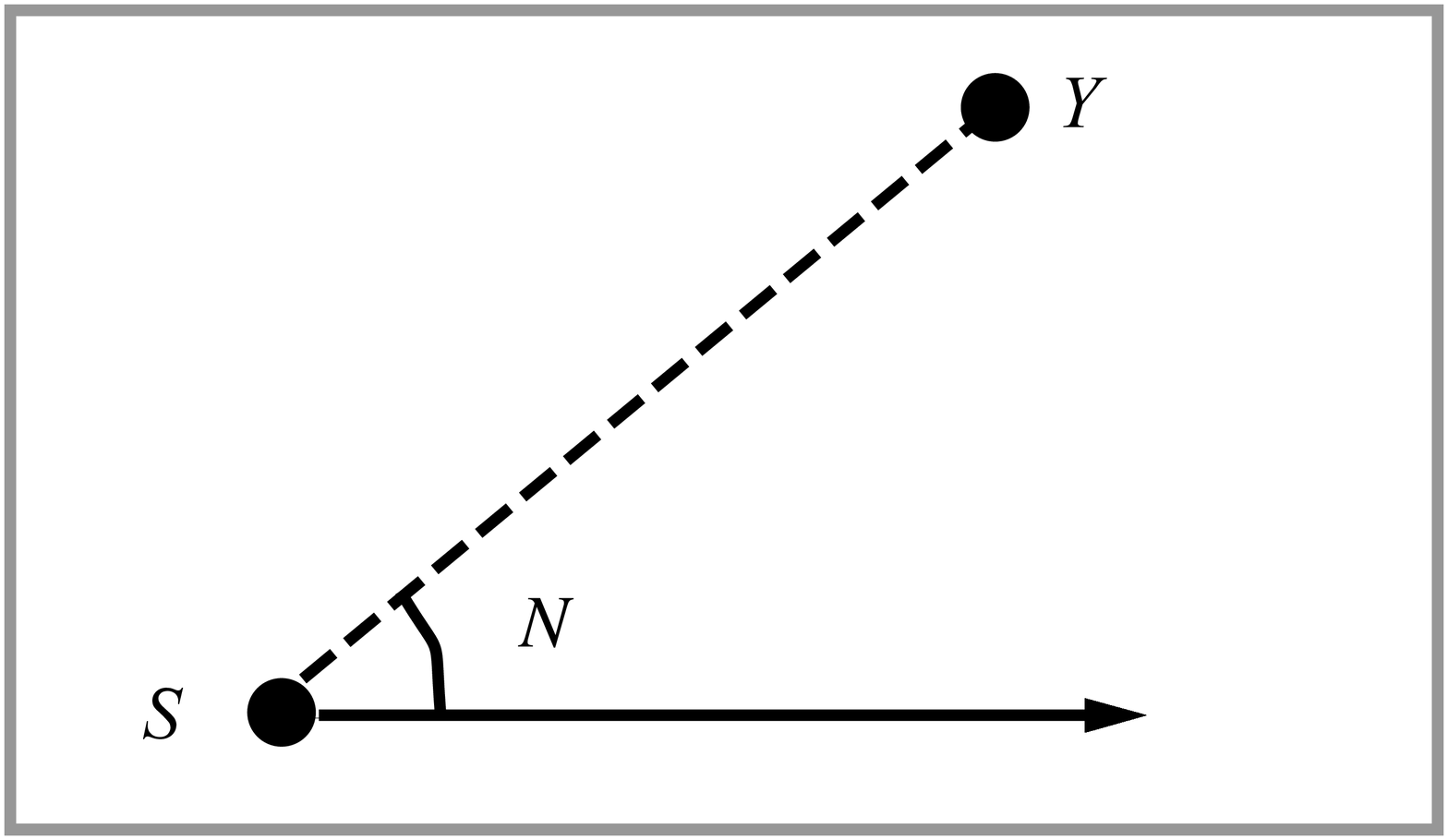}}
  \centerline{(a)}\medskip
\end{minipage}

\begin{minipage}[b]{1.0\linewidth}
  \centering
  \centerline{\includegraphics[width=6cm]{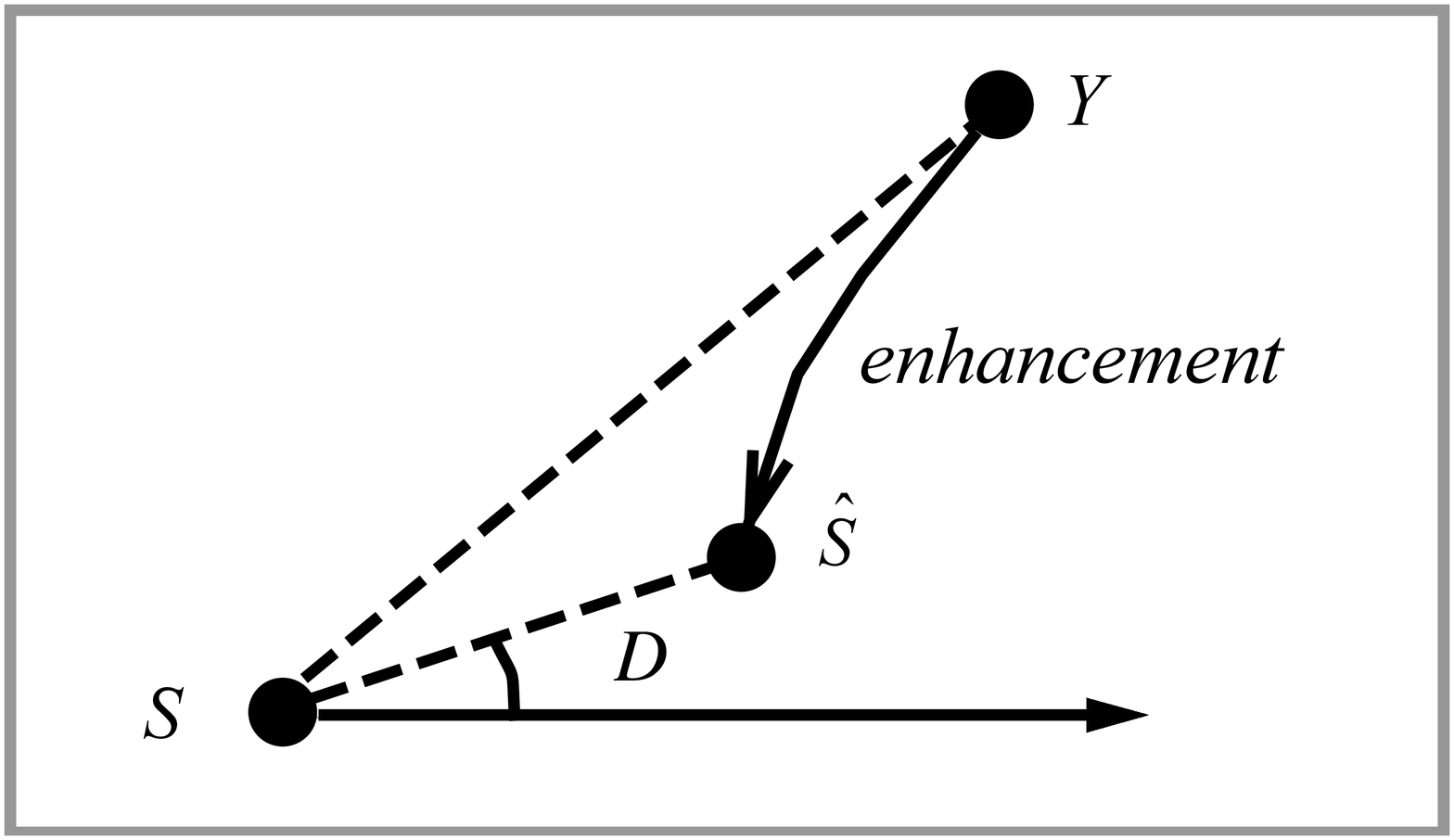}}
  \centerline{(b)}\medskip
\end{minipage}

\centering
 \caption{Illustration of the signal distortion problem. (a) The polar coordinate system. (b) Clean, noisy, and enhanced speech.}

\label{fig:distortion}

\end{figure}

Fig. \ref{fig:distortion} shows the deviation of $D$ from $N$ in an intuitive way. In this figure, spectral representations of different signals are plotted in a polar coordinate system. The center of the coordinate system denotes clean speech $S$. The distance between clean speech $S$ and noisy speech $Y$ indicates the intensity of noise, and the angle between $SY$ and a predetermined axis indicates noise type $N$. Fig. \ref{fig:distortion}(a) shows $S$ and its mixture with $N$, and Fig. \ref{fig:distortion}(b) illustrates the relative positions of $Y$ and enhanced speech $\hat{S}$.

\begin{figure*}[thb]

\centering
\includegraphics[width=17.6cm]{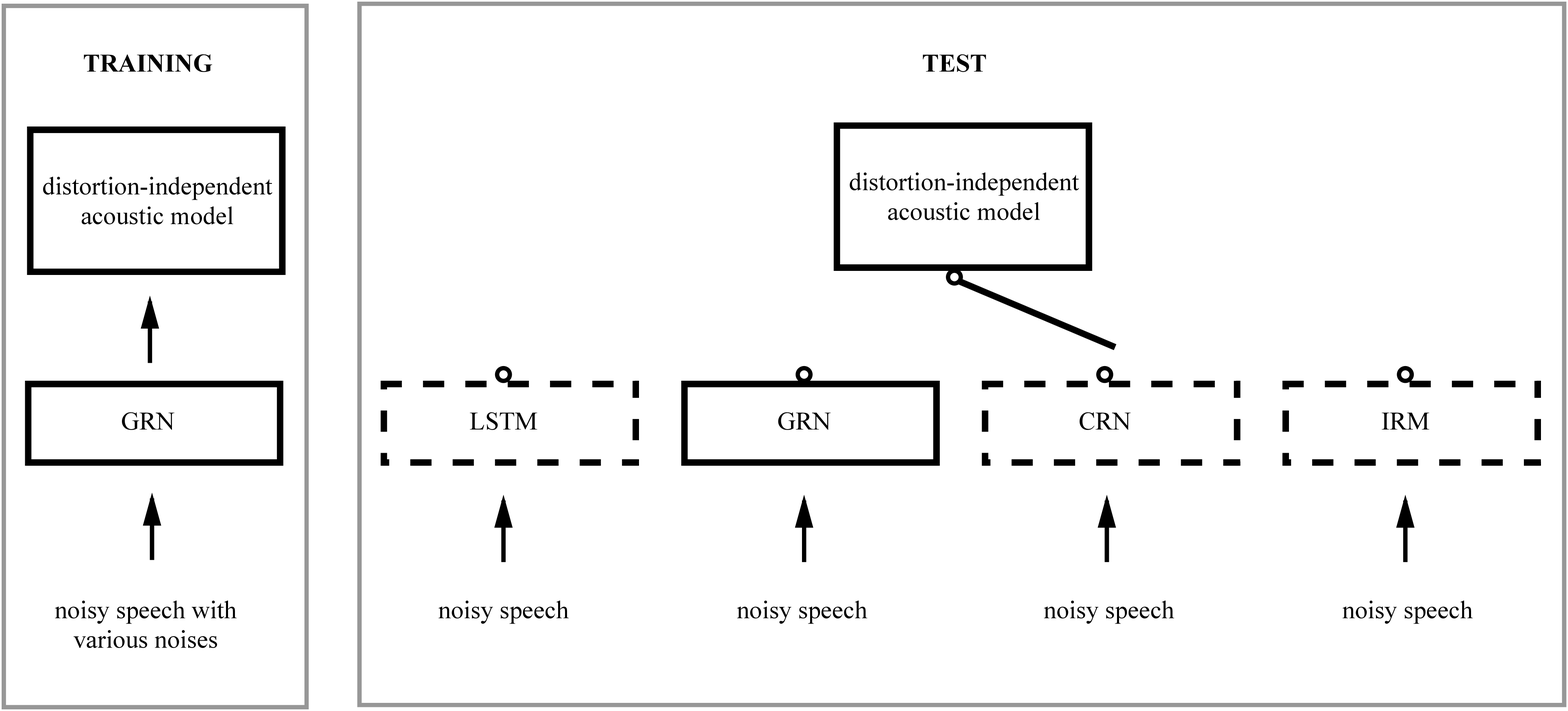}
\caption{Illustration of distortion-independent acoustic modeling. See text for the meaning of acronyms.}
\label{fig:frontends}

\end{figure*}

As is shown in Fig. \ref{fig:distortion}(b), compared with $Y$, $\hat{S}$ is typically closer to $S$. This corresponds to the observation that the SNR of enhanced speech is typically higher than that of noisy speech. In fact, many enhancement models are explicitly designed to elevate the SNR.

Along with the shorter distance to $S$, enhanced speech $\hat{S}$ may deviate from line $SY$. Such a noise mismatch between $Y$ and $\hat{S}$ may degrade the performance of ASR systems trained only on $Y$. This may be the main cause of the distortion problem. In fact, for two utterances mixed with the same kind of noise at different SNRs, experimental results suggest that the one with higher SNR typically yield higher recognition performance. Note that, because of the similarity of masking-based and mapping-based speech enhancement in terms of distortion, the analysis above is expected to be valid for mapping-based systems as well.

\subsection{Distortion-Independent Acoustic Modeling}
For ASR backends trained on noisy speech and evaluated on enhanced speech, the input data for training and evaluation can be expressed as (\ref{eq:train}) and (\ref{eq:eval}), respectively,

\begin{equation}
\label{eq:train}
|Y_{tr}| = |S_{tr} + N_{tr}|
\end{equation}

\begin{equation}
\label{eq:eval}
|\hat{S}_{eval}| = |S_{eval} + D_{eval}| 
\end{equation}
where $D_{eval} = N_{eval} \otimes M_{eval} - S_{eval} \otimes \overline{M_{eval}}$. Subscripts $tr$ and $eval$ denote training and evaluation, respectively. $Y_{tr}$, $S_{tr}$, and $N_{tr}$ are the spectral representations of noisy speech, clean speech, and additive noise in training, respectively. $\hat{S}_{eval}$ is the enhanced speech in evaluation. $D_{eval}$ denotes the distortion in enhanced speech and $M_{eval}$ the T-F mask in evaluation.

Based on our analysis in the previous subsection, the mismatch between $N_{tr}$ and $D_{eval}$ is the cause of the distortion problem. In speech recognition corpora such as Aurora and CHiME series, only a limited number of noises are provided for training. In addition, noise types are shared between training and evaluation on these corpora. ASR systems trained with such noisy speech may not perform well on enhanced speech, which contains mismatched interference $D_{eval}$. Moreover, same noise types between training and evaluation give an advantage to unenhanced evaluation data. This is likely a main reason why speech enhancement does not improve recognition performance on these tasks.

To alleviate the distortion problem, $N_{tr}$ can be modified in two ways. If we view $D_{eval}$ as a special type of additive noise, a straightforward way is to increase the scope of $N_{tr}$. Since this strategy typically uses a large variety of additive noises to train acoustic models, we denote it noise-independent acoustic modeling. An advantage of noise-independent training is that its efficacy is not influenced by speech enhancement frontends. This acoustic modeling strategy, however, does not account for the fact that additive noises may differ significantly from distortions. Another strategy to alleviate the distortion problem is to train the acoustic model directly with enhanced speech, i.e.

\begin{equation}
\label{eq:tr_enh}
|\hat{S}_{tr}| = |S_{tr} + D_{tr}|
\end{equation}
where $\hat{S}_{tr}$ denotes enhanced training speech and $D_{tr}$ refers to the distortion in it.

We investigate a distortion-independent acoustic modeling method based on (\ref{eq:tr_enh}). The training set consists of a large variety of enhanced speech generated by a single well-trained speech enhancement frontend. The input to the speech enhancement model is noisy speech with various types of additive noise. An advantage of distortion-independent acoustic modeling is that $D_{tr}$ in enhanced training speech is similar to $D_{eval}$ during evaluation. The main concern is its generalization ability to other speech enhancement frontends. Since most supervised speech enhancement models can be viewed as nonlinear mapping from noisy speech to clean speech, distortion-independent acoustic model may be able to work with speech enhancement frontends different from the frontend used for training.

Fig. \ref{fig:frontends} illustrates distortion-independent acoustic modeling. The left diagram depicts the training stage and the right one testing. In the right diagram, speech enhancement blocks with dashed lines denote those not used during training. In this study, we evaluate three existing speech enhancement models: gated residual network (GRN) \cite{tan2018gated}, LSTM \cite{chen2017long}, and convolutional recurrent network (CRN) \cite{tan2018convolutional}. We also add the IRM as another enhancement frontend. The switch in the right diagram denotes the coupling between a distortion-independent acoustic model and various enhancement frontends.

\subsection{Types of Acoustic Models}
In addition to noise-independent and distortion-independent acoustic models, we investigate three other types of acoustic models: clean, noise-dependent, and noise-mismatched. The clean acoustic model is trained using clean speech. In corpora containing both additive noise and reverberation, clean refers to reverberant speech without noise. The noise-dependent acoustic model is trained using only one type of noise and is tested on the same type of noise. This experimental setup represents typical robust speech recognition evaluations. The noise-mismatched acoustic model also uses a single type of noise during training, but it is tested on noises different from those for training.

\subsection{Utterance-Wise Recurrent Dropout for Acoustic Model Training}
Utterance-wise recurrent dropout has been shown to be effective for acoustic model training on the CHiME-4 corpus \cite{wang2018utterance}. A typical LSTM layer is described in Eqs. (\ref{eq:lstm1}), (\ref{eq:lstm2}), and (\ref{eq:lstm3}) below:

\begin{equation}
\label{eq:lstm1}
\begin{pmatrix}
\textbf{i}_t\\
\textbf{f}_t\\
\textbf{o}_t\\
\textbf{g}_t\\
\end{pmatrix}
=
\begin{pmatrix}
\sigma(\textbf{W}_{i}\textbf{x}_t+\textbf{U}_{i}\textbf{h}_{t-1}+\textbf{b}_i)\\
\sigma(\textbf{W}_{f}\textbf{x}_t+\textbf{U}_{f}\textbf{h}_{t-1}+\textbf{b}_f)\\
\sigma(\textbf{W}_{o}\textbf{x}_t+\textbf{U}_{o}\textbf{h}_{t-1}+\textbf{b}_o)\\
f(\textbf{W}_{g}\textbf{x}_t+\textbf{U}_{g}\textbf{h}_{t-1}+\textbf{b}_g)
\end{pmatrix}
\end{equation}

\begin{equation}
\label{eq:lstm2}
\textbf{c}_t=\textbf{f}_t\otimes\textbf{c}_{t-1}+\textbf{i}_t\otimes\textbf{g}_t
\end{equation}

\begin{equation}
\label{eq:lstm3}
\textbf{h}_t=\textbf{o}_t\otimes f(\textbf{c}_t)
\end{equation}
The dropout method can be expressed as follows:

\begin{equation}
\label{eq:dropout}
\begin{pmatrix}
\textbf{i}_t\\
\textbf{f}_t\\
\textbf{o}_t\\
\textbf{g}_t\\
\end{pmatrix}
=
\begin{pmatrix}
\sigma(\textbf{W}_{i}d_{xit}(\textbf{x}_t)+\textbf{U}_{i}d_{hi}(\textbf{h}_{t-1}) +\textbf{b}_i)\\
\sigma(\textbf{W}_{f}d_{xft}(\textbf{x}_t)+\textbf{U}_{f}d_{hf}(\textbf{h}_{t-1}) +\textbf{b}_f)\\
\sigma(\textbf{W}_{o}d_{xot}(\textbf{x}_t)+\textbf{U}_{o}d_{ho}(\textbf{h}_{t-1}) +\textbf{b}_o)\\
f(\textbf{W}_{g}d_{xgt}(\textbf{x}_t)+\textbf{U}_{g}d_{hg}(\textbf{h}_{t-1}) +\textbf{b}_g)
\end{pmatrix}
\end{equation}
where $\textbf{i}_t$, $\textbf{f}_t$, and $\textbf{o}_t$ are the input, forget, and output gates at step $t$; $\textbf{g}_t$ is the vector of cell updates and $\textbf{c}_t$ denotes updated cell vector; $\textbf{c}_t$ is used to update hidden state $\textbf{h}_t$; $\sigma$ is a sigmoid function and $f$ is typically chosen to be $tanh$. $\textbf{W}$ and $\textbf{U}$ are the weight matrices for the input vector $\textbf{x}_t$ and hidden vector $\textbf{h}_{t-1}$, respectively. $\textbf{b}$ denotes the bias term. The dropout function is denoted as $d()$. Subscripts $x$ and $h$ refer to the two corresponding feature vectors and $i$, $f$, $o$, $g$ correspond to the four LSTM components. Dropout functions with subscript $t$ are conventional frame-wise dropout, and those without $t$ are recurrent, i.e. they use the same dropout mask at different time steps.

Utterance-wise recurrent dropout is designed to be both recurrent and have little temporal information loss. Four independently sampled utterance-wise masks are applied to $\textbf{h}_{t-1}$. For the dropout on $\textbf{x}_t$, we opt for a conventional frame-wise method since utterance-wise dropout may completely lose the information in some feature dimensions.

\section{Experimental Setup}
\subsection{Datasets}
We use two corpora in our experiments. One of them is designed specially for this study and the other one follows the official CHiME-2 recipe.

\subsubsection{WSJ}
We compose a corpus by mixing clean speech in WSJ with additive noise. Although such simulated corpora are not commonly used in speech recognition, they are common in speech enhancement \cite{chen2016large,tan2018gated}.

Training sets for the five acoustic models are designed in the following way. For the clean acoustic model, the clean utterances in the original WSJ corpus are used directly. The noise-dependent acoustic model has two instances, each corresponding to a different noise. The noise-mismatched acoustic model also has two instances, but it differs from the noise-dependent acoustic model in that its training and testing noises are mismatched. The training sets for the noise-dependent and noise-mismatched acoustic models are the same. It contains 7138 utterances generated by mixing clean utterances with a training noise (ADTbabble or ADTcafeteria1) at SNRs randomly chosen from \{9dB, 6dB, 3dB, 0dB, -3dB, -6dB\}. ADTbabble and ADTcafeteria1 (available at http://www.auditec.com) are commonly used in speech enhancement tasks \cite{chen2016large,tan2018gated}. For the noise-independent acoustic model, the training set is generated by adding noise segments from a 10000 noise database (available at https://www.soundideas.com) to clean utterances at SNRs randomly chosen from the above six levels. The size of the noise-independent training set is 157036, 22 times that of the clean training set. The distortion-independent acoustic model is trained using GRN enhanced speech. GRN takes as input the noisy speech used for noise-independent acoustic model training. The distortion-independent training set thus also contains 157036 utterances.

A validation set is shared among the five acoustic models. It contains 1206 clean utterances from 10 speakers different from those used in training sets. Note that clean utterances are used directly in the validation set, avoiding biases to any specific noise.

The five acoustic models also share the same test set. It consists of 330 noisy utterances for each of the two test noises (ADTbabble and ADTcafeteria1) and at each of the six SNRs (i.e. \{9dB, 6dB, 3dB, 0dB, -3dB, -6dB\}). The total number of utterances is 3960. These utterances are from 12 speakers different from those in the training and validation sets.

Note that although ASR backends and enhancement frontends both use the 10k noise database, their actual noise segments are different. First, ASR backends only use the first halves of noises, and enhancement frontends the second halves. Second, noise segments are randomly selected for recognition and enhancement.

\subsubsection{CHiME-2}
CHiME-2 is a commonly used corpus for robust speech recognition. Different from WSJ, utterances in CHiME-2 contain room reverberation. We treat reverberant speech in CHiME-2 as clean speech.

Training sets for the five acoustic models are designed based on the official recipe of the CHiME-2 challenge. The reverberant acoustic model is trained using reverberant utterances. Since each recording in CHiME-2 has two channels, we apply an average operation to get the corresponding monaural utterance. The noise-dependent acoustic model exactly follows the CHiME-2 recipe. The noise-mismatched acoustic model tests the noise-dependent acoustic model on ADT noises (ADTbabble and ADTcafeteria1) rather than the CHiME-2 noises. Due to the limited number of noises provided in the CHiME-2 corpus, the noise-independent acoustic model is trained with additional noises from the 10k noise database. We mix reverberant utterances with noise segments. The SNR levels are the same as those for WSJ. The noise-independent training set contains 157036 utterances in total. For the distortion-independent acoustic model, the training set consists of 157036 utterances enhanced by GRN.

For the noise-dependent acoustic model, we apply a validation set consisting of noisy utterances. For the other four acoustic models, we use reverberant utterances.

In addition to the official CHiME-2 test set, we generate two other test sets containing ADT noises. The average results on ADT noises are reported in this paper.

Due to reverberation, speech enhancement models for CHiME-2 are trained to map from reverberant-noisy speech to reverberant speech. The training data for speech enhancement models are generated similarly to those for the noise-independent acoustic model.

\subsection{Implementation Details}
We use a wide residual bidirectional LSTM network (WRBN) as the DNN architecture of acoustic models \cite{jahn2016wide,wang2018utterance,wang2018filter}. For speech enhancement frontends, we adopt three models as illustrated in Fig. \ref{fig:frontends}. GRN is the main frontend in our experiments and is used to generates both training data and test data. Two other speech enhancement models, LSTM and CRN, generate additional test data for distortion-independent acoustic modeling. These three frontends use different training targets. GRN applies the phase sensitive mask (PSM), LSTM uses the IRM, and CRN is mapping based.

We couple enhancement frontends and ASR backends with enhanced waveforms. The preprocessing steps for the enhancement frontends include windowing and Fourier transform. We apply the Hamming window with window width 20ms and shift 10ms. The windowed waveform signals are then converted to 320-dimensional short-time Fourier transform (STFT) features. Speech enhancement models take as input the STFT magnitudes and generate masks or enhanced magnitudes. We combine enhanced magnitudes with the phase of noisy speech to resynthesize enhanced waveform signals. As for the feature preprocessing for ASR backends, we make modifications to the recipe in Kaldi and our previous experiments \cite{wang2018utterance,wang2018filter}. In order to avoid manually added interferences to enhanced speech, we skip most of preprocessing steps, including pre-emphasizing, dithering, and direct currency offset removal. Similar to speech enhancement frontends, we extract spectral features from enhanced waveform signals by applying the Hamming window and performing STFT. One difference is that STFT features for speech recognition have 512 dimensions. We then apply Mel filters to STFT magnitudes to generate Mel frequency features. The dimension of Mel features is 80. In order to avoid underflow, we add a small value $e^{-40}$ to Mel features and apply logarithm to the summation. The delta and delta-delta of log-Mel features are then generated, tripling the size of feature dimensionality. We calculate the mean value along time for each utterance and subtract it from the features. ASR backends take as input the normalized features and generate log posterior probabilities for senones. There are 1965 senones in our experiments. Subtracting log priors from log posteriors, we feed log likelihoods to the decoder in CHiME-2 to generate transcriptions.

In training the noise-independent and distortion-independent acoustic models, we monitor validation results after every 7138 utterances. This technique is commonly used in speech and language processing experiments.

\subsubsection{WSJ}
During training, most hyper-parameters for the five acoustic models are the same. The optimizer is Adam and dropout rate is 0.2. Initial learning rate is set to $10^{-3}$ for all acoustic models.

\subsubsection{CHiME-2}
For experiments on CHiME-2, the noise-mismatched acoustic model uses the well-trained noise-dependent acoustic model. Therefore, there are only four acoustic models on this corpus. The optimizer and dropout rate are the same as those on WSJ. The four acoustic models share the same initial learning rate of $10^{-4}$. 

\section{Evaluation Results and Analysis}
This section presents and analyzes our evaluation results on the five acoustic models. The results are provided separately for the WSJ and CHiME-2 corpora.

\subsection{Results on WSJ}
 \begin{table*}[ht]
 \setlength{\tabcolsep}{3pt} 
  \renewcommand{\arraystretch}{2} 
  \caption{WERs of the five acoustic models on WSJ. \emph{bab} and \emph{caf} denote ADTbabble and ADTcafeteria1, respectively. \emph{w/o} refers to noisy evaluation data without speech enhancement (i.e. unenhanced speech), and \emph{w/} evaluation data with enhancement.}
  \centering
    \begin{tabular}{| c | c  c | c  c | c  c | c  c | c  c | c  c | c  c | c  c | c c | c c |}
      \hline
      \multirow{3}{*}{SNR} & \multicolumn{4}{c|}{clean} & \multicolumn{4}{c|}{noise-dependent} & \multicolumn{4}{c|}{noise-mismatched} & \multicolumn{4}{c|}{noise-independent} & \multicolumn{4}{c|}{distortion-independent} \\
     \hhline{~--------------------}
     & \multicolumn{2}{c|}{bab} & \multicolumn{2}{c|}{caf} & \multicolumn{2}{c|}{bab} & \multicolumn{2}{c|}{caf} & \multicolumn{2}{c|}{bab} & \multicolumn{2}{c|}{caf} & \multicolumn{2}{c|}{bab} & \multicolumn{2}{c|}{caf} & \multicolumn{2}{c|}{bab} & \multicolumn{2}{c|}{caf} \\
     \hhline{~--------------------}
     & w/o & w/ & w/o & w/ & w/o & w/ & w/o & w/ & w/o & w/ & w/o & w/ & w/o & w/ & w/o & w/ & w/o & w/ & w/o & w/ \\
\hline
     9dB & 11.92 & \textbf{3.08} & 12.83 & \textbf{3.53} & \textbf{3.62} & 4.35 & \textbf{4.28} & 5.04 & 6.31 & \textbf{5.01} & 4.95 & \textbf{4.05} & 4.89 & \textbf{4.00} & 4.97 & \textbf{4.04} & 4.18 & \textbf{3.10} & 3.81 & \textbf{3.29} \\
     6dB & 22.19 & \textbf{4.11} & 22.06 & \textbf{6.15} & \textbf{4.28} & 5.04 & \textbf{5.55} & 6.39 & 9.83 & \textbf{5.94} & 7.98 & \textbf{5.77} & 7.14 & \textbf{4.86} & 7.17 & \textbf{5.55} & 5.10 & \textbf{4.00} & 5.59 & \textbf{4.80} \\
     3dB & 38.26 & \textbf{6.67} & 38.88 & \textbf{9.15} & \textbf{5.12} & 6.31 & \textbf{8.11} & 8.93 & 17.07 & \textbf{7.85} & 14.16 & \textbf{8.59} & 10.59 & \textbf{6.65} & 11.06 & \textbf{8.09} & 7.17 & \textbf{5.23} & 8.85 & \textbf{7.08} \\
     0dB & 60.32 & \textbf{12.46} & 58.25 & \textbf{17.34} & \textbf{7.55} & 9.64 & \textbf{12.07} & 14.68 & 28.41 & \textbf{11.68} & 26.13 & \textbf{14.05} & 18.23 & \textbf{10.74} & 17.56 & \textbf{14.18} & 12.87 & \textbf{9.19} & 15.21 & \textbf{12.85} \\
     -3dB & 82.44 & \textbf{23.24} & 79.15 & \textbf{32.51} & \textbf{12.55} & 18.03 & \textbf{21.93} & 27.72 & 46.16 & \textbf{21.39} & 48.48 & \textbf{26.43} & 31.89 & \textbf{19.71} & 31.14 & \textbf{26.64} & 24.30 & \textbf{17.13} & 27.82 & \textbf{24.58} \\
     -6dB & 93.16 & \textbf{44.76} & 91.44 & \textbf{56.25} & \textbf{22.66} & 34.34 & \textbf{40.13} & 48.40 & 71.64 & \textbf{38.05} & 74.67 & \textbf{49.52} & 54.06 & \textbf{36.19} & 53.99 & \textbf{47.94} & 45.41 & \textbf{33.55} & 50.68 & \textbf{45.17} \\
     \hline
     avg & 51.4 & {\textbf{15.7}} & 50.4 & {\textbf{20.8}} & {\textbf{9.3}} & 13.0 & {\textbf{15.3}} & 18.5 & 29.9 & {\textbf{15.0}} & 29.4 & {\textbf{18.1}} & 21.1 & {\textbf{13.7}} & 21.0 & {\textbf{17.7}} & 16.5 & \textbf{12.0} & 19.4 & \textbf{16.3} \\
\hline
  \end{tabular}
 \label{tb:wsj_main}
  \end{table*}

TABLE \ref{tb:wsj_main} shows the WERs of the five acoustic models on WSJ. We use ADTbabble and ADTcafeteria1 as the noises for evaluation. The clean acoustic model clearly benefits from speech enhancement, as enhanced speech has a higher SNR than the corresponding noisy speech. 

For the noise-dependent acoustic model, consistent with previous observations \cite{narayanan2014investigation,du2014robust,jahn2016wide}, the results on unenhanced speech are better. Based on our analysis on the distortion problem, the performance degradation on enhanced speech is caused by the mismatch between $N_{tr}$ and $D_{eval}$.

The noise-mismatched acoustic models are able to benefit from speech enhancement in our experiments on WSJ. Such an ability, however, is influenced by the type of noise used for testing. We will discuss this more after presenting the results on CHiME-2. In TABLE \ref{tb:wsj_main}, we observe that the results of noise-dependent acoustic models are much better than those of noise-mismatched acoustic models on unenhanced speech. This indicates that acoustic models trained on one noise cannot generalize to untrained noises. This performance degradation caused by noise mismatch supports our analysis on the distortion problem.

The noise-independent acoustic model also benefits from speech enhancement on WSJ. This indicates that 10k additive noises can capture a lot of the distortions on WSJ. The efficacy of noise-independent acoustic modeling, however, may be influenced by factors such as reverberation, as will be shown in the results on the CHiME-2 corpus.

For the distortion-independent acoustic model, the results on enhanced speech are better than those on unenhanced speech. This shows that distortion-independent acoustic models are able to alleviate the distortion problem caused by GRN. Moreover, the results of the distortion-independent acoustic model are better than those of the noise-independent model. Note that both noise-independent and distortion-independent acoustic models are tested on noises different from those used during training. The strong performance of our distortion-independent acoustic model shows that large-scale training with various distortions generalizes well to untrained distortions.

Along each column of TABLE \ref{tb:wsj_main}, there is a clear performance degradation as SNR reduces, consistent with our analysis on the cause of the distortion problem.

\begin{table*}[ht]
 \setlength{\tabcolsep}{6pt} 
  \renewcommand{\arraystretch}{1.7} 
  \caption{WERs of the distortion-independent acoustic model on WSJ with other frontends. See TABLE \ref{tb:wsj_main} caption for notations.}
  \centering
    \begin{tabular}{| c | c c | c c | c c | c c |}
      \hline
	 \multirow{2}{*}{SNR} & \multicolumn{2}{c|}{unenhanced} & \multicolumn{2}{c|}{LSTM} & \multicolumn{2}{c|}{CRN} & \multicolumn{2}{c|}{IRM} \\
     \hhline{~--------}
	& bab & caf & bab & caf & bab & caf & bab & caf\\
\hline
     9dB  & 4.18 & 3.81 & 3.36 & 3.21 & 3.27 & 4.09 & 2.73 & 2.73\\
     6dB & 5.10 & 5.59 & 4.24 & 4.88 & 4.13 & 4.65 & 2.75 & 2.88\\
     3dB & 7.17 & 8.85 & 5.70 & 7.64 & 5.03 & 7.53 & 2.84 & 2.65\\
     0dB & 12.87 & 15.21 & 9.02 & 13.17 & 8.39 & 11.53 & 2.88 & 2.86\\
     -3dB & 24.30 & 27.82 & 17.95 & 25.14 & 14.37 & 22.38 & 3.05 & 2.76\\
     -6dB & 45.41 & 50.68 & 34.99 & 47.80 & 28.19 & 40.82 & 2.91 & 2.93\\
     \hline
     avg & 16.5 & 19.4 & 12.5 & 17.0 & 10.6 & 15.2 & 2.9 & 2.8\\
\hline
  \end{tabular}
  \label{tb:wsj_general}
  \end{table*}

In TABLE \ref{tb:wsj_general}, we present the results of the distortion-independent acoustic model when coupled with speech enhancement frontends different from the one used during training. From the table, we observe that both LSTM and CRN yield better results than unenhanced speech. This shows that the distortion-independent acoustic model is able to generalize to different enhancement frontends. This also suggests that there may be a common pattern in the distortions introduced by supervised speech enhancement models. 

Comparing the results of LSTM, CRN, and IRM, we find that for the distortion-independent acoustic model, the improvement of speech enhancement quality results in the improvement of recognition performance. In real-world applications, this suggests that a distortion-independent model need not to be retrained when a more advanced speech enhancement frontend is applied. In addition, the distortion-independent acoustic model on WSJ may be used to provide an indicator on the modeling ability of different speech enhancement frontends. Note that at different SNRs, the IRM results vary slightly, which may be due to the waveform resynthesis during speech enhancement. When the distortion-independent acoustic model is evaluated on clean speech, the average WER is 2.7\%. For the clean acoustic model evaluated on clean speech, the WER is 2.0\%.

\subsection{Results on CHiME-2}
  \begin{table*}[ht]
 \setlength{\tabcolsep}{3pt} 
  \renewcommand{\arraystretch}{2} 
  \caption{WERs of the five acoustic models on CHiME-2. \emph{chime-2} denotes the official CHiME-2 evaluation set. \emph{ADT} refers to the average WER of ADTbabble and ADTcafeteria1. See TABLE \ref{tb:wsj_main} caption for other notations.}
  \centering
    \begin{tabular}{| c | c c | c c | c c | c c | c c | c c | c c | c c |}
      \hline
     \multirow{3}{*}{SNR} & \multicolumn{4}{c|}{reverberant} & \multicolumn{2}{c|}{noise-dependent} & \multicolumn{2}{c|}{noise-mismatched} & \multicolumn{4}{c|}{noise-independent} & \multicolumn{4}{c|}{distortion-independent}\\
     \hhline{~----------------}
     & \multicolumn{2}{c|}{chime-2} & \multicolumn{2}{c|}{ADT} & \multicolumn{2}{c|}{chime-2} & \multicolumn{2}{c|}{ADT} & \multicolumn{2}{c|}{chime-2} & \multicolumn{2}{c|}{ADT} & \multicolumn{2}{c|}{chime-2} & \multicolumn{2}{c|}{ADT}\\
     \hhline{~----------------}
     & w/o & w/ & w/o & w/ & w/o & w/ & w/o & w/ & w/o & w/ & w/o & w/ & w/o & w/ & w/o & w/ \\
\hline
9dB & 31.27 & \textbf{10.50} & 31.03 & \textbf{11.40} & \textbf{5.49} & 5.81 & \textbf{7.82} & 8.33 & 6.63 & \textbf{6.37} & \textbf{6.59} & 7.98 & 7.42 & \textbf{5.51} & 10.20 & \textbf{6.60}\\
6dB & 38.69 & \textbf{13.67} & 47.53 & \textbf{19.00} & \textbf{6.26} & 7.98 & \textbf{10.28} & 11.76 & \textbf{7.72} & 7.92 & \textbf{8.66} & 11.04 & 8.61 & \textbf{6.54} & 13.27 & \textbf{8.64}\\
3dB & 46.85 & \textbf{17.26} & 67.50 & \textbf{31.68} & \textbf{6.78} & 8.33 & \textbf{18.03} & 20.46 & 8.82 & \textbf{8.78} & \textbf{14.00} & 19.35 & 10.01 & \textbf{7.10} & 20.99 & \textbf{14.73}\\
0dB & 57.33 & \textbf{23.73} & 85.96 & \textbf{50.89} & \textbf{8.95} & 11.26 & \textbf{30.07} & 34.38 & \textbf{10.69} & 11.62 & \textbf{23.72} & 30.99 & 12.93 & \textbf{9.70} & 32.30 & \textbf{22.76}\\
-3dB & 62.94 & \textbf{29.91} & 93.49 & \textbf{71.89} & \textbf{9.98} & 14.48 & \textbf{50.34} & 55.30 & \textbf{13.06} & 13.30 & \textbf{39.04} & 51.60 & 14.85 & \textbf{11.04} & 51.02 & \textbf{37.74}\\
-6dB & 72.31 & \textbf{39.87} & 95.58 & \textbf{88.79} & \textbf{14.83} & 19.05 & \textbf{75.65} & 78.83 & \textbf{17.45} & 19.80 & \textbf{60.73} & 76.08 & 21.80 & \textbf{15.45} & 75.54 & \textbf{58.24}\\
\hline
avg & 51.6 & \textbf{22.5} & 70.2 & \textbf{45.6} & \textbf{8.7} & 11.2 & \textbf{32.0} & 34.8 & \textbf{10.7} & 11.3 & \textbf{25.5} & 32.8 & 12.6 & \textbf{9.2} & 33.9 & \textbf{24.8}\\
\hline
  \end{tabular}
  \label{tb:chime2_main}
  \end{table*}

TABLE \ref{tb:chime2_main} presents the WERs of the five acoustic models on CHiME-2. The noises used for evaluation include chime-2 noises and ADT. WERs on ADT are the averages of those on ADTbabble and ADTcafeteria1. The reverberant acoustic model on CHiME-2 corresponds to the clean acoustic model on WSJ. It is clear that the reverberant acoustic model benefits from speech enhancement.

The noise-dependent acoustic model follows the official training recipe of the CHiME-2 challenge. Similar to prior observations \cite{narayanan2013ideal,du2014robust}, the noise-dependent acoustic model does not benefit from speech enhancement, which is in line with the results in TABLE \ref{tb:wsj_main}.

We test the noise-mismatched acoustic model on ADT noises. Different from the results on WSJ, the noise-mismatched acoustic model does not perform better on enhanced speech. Note that the experiments on CHiME-2 use CHiME-2 noises for training, whereas the experiments on WSJ use ADT noises. The inconsistent results on the two corpora indicate that the ability of the noise-mismatched acoustic model to overcome the distortion problem may depend on the noise used for testing.

The noise-independent acoustic model on CHiME-2 does not gain performance improvement on enhanced speech. This is again different from the corresponding results on WSJ. On the CHiME-2 corpus, room impulse responses (RIRs) are different between training and testing \cite{vincent2013second}. Although we use a large variety of additive noises to train the noise-independent acoustic model, the RIR mismatch still exists. During testing, distortions introduced by speech enhancement thus deviate from the 10k additive noises used for training. Note that at SNR level 9dB and 3dB, enhanced speech performs better than unenhanced speech on the CHiME-2 corpus.

The distortion-independent acoustic model is able to benefit from speech enhancement. On both CHiME-2 and ADT noises, distortion-independent acoustic model outperforms noise-independent acoustic model. The ability of distortion-independent acoustic modeling to benefit from speech enhancement shows that large-scale training on a variety of distortions generalizes to untrained distortions.

  \begin{table*}[ht]
 \setlength{\tabcolsep}{6pt} 
  \renewcommand{\arraystretch}{1.7} 
  \caption{WERs of the distortion-independent acoustic model on CHiME-2 with other frontends. See TABLE \ref{tb:chime2_main} caption for notations.}
  \centering
    \begin{tabular}{| c | c c | c c | c  c | c c |}
      \hline
   \multirow{2}{*}{SNR} & \multicolumn{2}{c|}{unenhanced} & \multicolumn{2}{c|}{LSTM} & \multicolumn{2}{c|}{CRN} & \multicolumn{2}{c|}{IRM}\\
     \hhline{~--------}
	& chime-2 & ADT & chime-2 & ADT & chime-2 & ADT & chime-2 & ADT\\
\hline
     9dB  & 7.42 & 10.20 & 5.79 & 7.61 & 6.65 & 7.50 & 3.40 & 3.66\\
     6dB & 8.61 & 13.27 & 7.47 & 10.21 & 7.68 & 10.09 & 3.44 & 3.64\\
     3dB & 10.01 & 20.99 & 8.63 & 17.57 & 9.04 & 15.57 & 3.34 & 3.62\\
     0dB & 12.93 & 32.30 & 11.36 & 28.47 & 11.25 & 25.73 & 3.38 & 3.73\\
     -3dB & 14.85 & 51.02 & 14.16 & 44.83 & 13.51 & 41.12 & 3.74 & 3.95\\
     -6dB & 21.80 & 75.54 & 19.41 & 67.08 & 18.06 & 62.33 & 3.31 & 4.10\\
     \hline
     avg & 12.6 & 33.9 & 11.1 & 29.3 & 11.0 & 27.1 & 3.4 & 3.8\\
\hline
  \end{tabular}
  \label{tb:chime2_general}
  \end{table*}
  
  TABLE \ref{tb:chime2_general} shows the results of the distortion-independent acoustic model when used with different speech enhancement frontends. Similar to the experiments on WSJ, distortion-independent acoustic model is tested on LSTM and CRN enhanced speech. The results of both LSTM and CRN enhanced speech are better than those of unenhanced speech. This indicates the ability of the distortion-independent acoustic model to work with various speech enhancement frontends. This also suggests that distortions introduced by different supervised enhancement models have certain similarities.

On IRM enhanced speech, the distortion-independent acoustic model performs very well. This suggests that as speech enhancement research progresses, speech recognition performances of the distortion-independent acoustic model should also improve. The average WER of the distortion-independent acoustic model on reverberant speech is 3.4\%. For the reverberant acoustic model evaluated on reverberant speech, the WER is 2.8\%.

  \begin{table*}[ht]
 \setlength{\tabcolsep}{3pt} 
  \renewcommand{\arraystretch}{2} 
  \caption{WER comparisons between the proposed models and prior work on CHiME-2.}
  \centering
    \begin{tabular}{| c | c c c c c c | c |}
    \hline
    models & 9dB & 6dB & 3dB & 0dB & -3dB & -6dB & avg \\
    \hline
    Wang and Wang \cite{wang2016joint} & 6.61 & 6.86 & 8.67 & 10.39 & 13.02 & 18.23 & 10.6\\
    \hline
    Plantinga \emph{et al.} \cite{plantinga2018exploration} & - & - & - & - & - & - & 9.3 \\
    \hline
    distortion-independent & 5.51 & 6.54 & 7.10 & 9.70 & 11.04 & 15.45 & 9.2 \\
    \hline
    noise-dependent & \textbf{5.49} & \textbf{6.26} & \textbf{6.78} & \textbf{8.95} & \textbf{9.98} & \textbf{14.83} & \textbf{8.7}\\
    \hline
  \end{tabular}
  \label{tb:chime2_comparison}
  \end{table*}

TABLE \ref{tb:chime2_comparison} shows a comparison of ASR systems in this study with those in prior work. It is worth noting that our distortion-independent acoustic model achieves a 9.2\% WER, which is better than the previous best systems on the CHiME-2 corpus \cite{plantinga2018exploration,wang2016joint}. For the noise-dependent acoustic model, we achieve an average WER of 8.7\%, outperforming the previous best system by 6.5\% relatively. Note that, in order to avoid the influence of model adaptation on our analysis of the distortion problem, we do not apply speaker adaptation to our models. The good results of our proposed models suggest that the observations in this study are likely valid for real world systems.

\section{Concluding Remarks}
The distortion problem occurs when we apply speech enhancement as a frontend for ASR tasks. This study treats the distortion problem as a noise mismatch between training and testing. We categorize acoustic models into five types and examine each of them for their ability to overcome the distortion problem. Distortion-independent acoustic modeling emerges as the best among the five acoustic models. Its ability to generalize to untrained noises suggests the utility of large-scale training for acoustic modeling. We also show that the distortion-independent acoustic model is able to work with various speech enhancement frontends. In addition, the WERs of our proposed distortion-independent and noise-dependent acoustic models both outperform the previous best system on the CHiME-2 corpus.

Future work on the distortion problem includes using ASR features as the training target of speech enhancement models, applying time-domain speech enhancement frontends, and investigating distortion-independent training for end-to-end ASR systems.

\newpage

\bibliographystyle{IEEEtranS}
\bibliography{mybib}



\vfill

\end{document}